\documentclass[conference]{IEEEtran}
\IEEEoverridecommandlockouts
\usepackage[utf8]{inputenc}
\usepackage{cite}
\usepackage{amsmath,amssymb,amsfonts}
\usepackage{algorithmic}
\usepackage{graphicx}
\usepackage{textcomp}
\usepackage{xcolor}
\usepackage{graphicx}
\usepackage[english]{babel}
\usepackage{booktabs}
\usepackage{url}
\def\BibTeX{{\rm B\kern-.05em{\sc i\kern-.025em b}\kern-.08em
    3T\kern-.1667em\lower.7ex\hbox{E}\kern-.125emX}}
\usepackage[colorinlistoftodos]{todonotes}

\definecolor{chestnut}{rgb}{0.8, 0.36, 0.36}

\definecolor{chestnut}{rgb}{0.8, 0.36, 0.36}
\begin{document}
\raggedbottom
\title{Linking Code and Documentation Churn: \\ Preliminary Analysis}

\author{\IEEEauthorblockN{Ani Hovhannisyan}
\IEEEauthorblockA{
\textit{NAIST, Japan}\\
hovhannisyan.ani.hb7@is.naist.jp
}

\and
\IEEEauthorblockN{Youmei Fan}
\IEEEauthorblockA{
\textit{NAIST, Japan}\\
fan.youmei.fs2@is.naist.jp
}

\and
\IEEEauthorblockN{Gema Rodríguez-Pérez}
\IEEEauthorblockA{
\textit{UBC, Canada} \\
gema.rodriguezperez@ubc.ca
}

\and
\IEEEauthorblockN{Raula Gaikovina Kula}
\IEEEauthorblockA{
\textit{Osaka University, Japan}\\
raula-k@is.naist.jp
}
}

\maketitle
\begin{abstract}
Code churn refers to the measure of the amount of code added, modified, or deleted in a project and is often used to assess codebase stability and maintainability. Program comprehension or how understandable the changes are, is equally important for maintainability. Documentation is crucial for knowledge transfer, especially when new maintainers take over abandoned code. We emphasize the need for corresponding documentation updates, as this reflects project health and trustworthiness as a third-party library. Therefore, we argue that every code change should prompt a documentation update (defined as documentation churn). Linking code churn changes with documentation updates is important for project sustainability, as it facilitates knowledge transfer and reduces the effort required for program comprehension. This study investigates the synchrony between code churn and documentation updates in three GitHub open-source projects. We will use qualitative analysis and repository mining to examine the alignment and correlation of code churn and documentation updates over time. We want to identify which code changes are likely synchronized with documentation and to what extent documentation can be auto-generated. Preliminary results indicate varying degrees of synchrony across projects, highlighting the importance of integrated concurrent documentation practices and providing insights into how recent technologies like AI, in the form of Large Language Models (i.e., LLMs), could be leveraged to keep code and documentation churn in sync. The novelty of this study lies in demonstrating how synchronizing code changes with documentation updates can improve the development lifecycle by enhancing diversity and efficiency.
\end{abstract}

\begin{IEEEkeywords}
code churn,
software maintenance,
knowledge transfer,
automated documentation generation
\end{IEEEkeywords}

\section{Introduction and Motivation}
\label{introduction}

In the realm of software development, maintaining up-to-date documentation alongside code changes is critical for project sustainability, future maintenance, and user convenience. Code churn, which refers to the changes made to the codebase, often includes significant updates that should ideally be mirrored in the software documentation. However, developers frequently overlook this synchrony, leading to discrepancies that can hamper project understanding and maintenance. This study explores the extent to which code churn changes align with documentation updates, focusing on three open-source projects hosted on GitHub. By addressing this gap, we aim to highlight the importance of integrated documentation practices for the efficacy and longevity of software projects.

Previous research has highlighted the challenges associated with keeping software documentation current with the evolving codebase. Studies have shown that inconsistencies between code and documentation can lead to increased maintenance costs and reduced software quality. In other cases, developers become less interested in difficult-to-maintain projects due to the lack of documentation and knowledge, eventually leading the project to its death \cite{gaikovina2023life}. 

Even though efforts were made towards automated documentation generation \cite{favre2012linking, rai22review, mcburney14auto} and tools for identifying outdated documentation, there still remains a need for comprehensive analysis of projects to understand the practical alignment between code churn and documentation updates over time. This study builds on existing work by applying empirical methods to evaluate this alignment in open-source projects, offering insights into current practices and potential improvements.

Moreover, creating thorough documentation every time a code change is submitted is crucial not only to enhance code comprehension and maintainability, but for fostering diversity and inclusion. Prior work reveals that diversity and inclusion enhance software development~\cite{shameer2023relationship}. However, documentation can pose significant challenges for developers, which can impact the efficacy and longevity of projects. Older developers often find documentation to be a major barrier when starting contributing \cite{davidson2014older}, women frequently join later than men to open source communities and in non-coding roles~\cite{robles2016women}. Helping to understand how to improve documentation, might contribute to avoiding the lack of diversity and inclusion in open source teams.

We can help to improve documentation by automating it using LLMs. They have been used to automate  software development processes already \cite{de2024fine,white2024chatgpt}. LLMs could alleviate the tedious tasks of writing documentation, making it easier for diverse individuals to focus on coding and other technical contributions. Also, the use of LLMs can break down language barriers, making writing documentation more accessible to non-native speakers. This  not only supports women and older adults in contributing more effectively, but also fosters a more inclusive and diverse environment. 
\section{Preliminary Empirical Study}
\label{prelimitary}

Our preliminary study examines three open-source GitHub projects: Apache HTTP Server \footnote{\url{https://github.com/apache/httpd}},
GNOME Gimp \footnote{\url{https://github.com/GNOME/gimp}},
 and GNOME GTK \footnote{\url{https://github.com/GNOME/gtk}}.
These were chosen for their large codebases, activity levels, and comprehensive documentation~\cite{robles2024role}. We collected data on code churn and documentation updates over a set period, with qualitative analysis assessing the nature and significance of these changes.

To investigate the synchrony between code churn and documentation updates, we utilized a two-pronged approach, combining quantitative and qualitative methods. First, repository mining was conducted to extract data on code changes and documentation for each commit from Git version control data. The quantitative data was then analysed to identify churn categories. Then, a qualitative review of selected churns was performed. 
Manual categorization of code churns into three categories, and a mix, was performed to understand the context of the changes within committed files. An example of file extensions is:

\begin{itemize}
    \item Source Code - \texttt{.c, .cpp, .h, .hpp, .js, .py}
    \item Documentation - \texttt{docs/*, .md, .rtf, .txt}
    \item Configuration - \texttt{.yml, .mk, build/*, conf/*}
    \item Mix - \texttt{.c, .h,  Makefile, Readme, NWGNU*}
\end{itemize}

Commits tha contain mixed code churns of Source Code and Documentation, or Source Code and Documentation and Configuration, were identified as Mix.

\begin{table}[t]
    \caption{Preliminary results showing Code Churn Categories.}
    \label{tab:repo_stats}
    \centering
    \begin{tabular}{lrrrr}
        \toprule
            Project  & Code & Documentation & Configuration & Mix\\
        \midrule 
            HTTP   & 14,768 & 2,595 & 9,313 & 5,417\\
             Gimp  & 22,317 & 1,893 & 1,799 & 21,467\\
              GTK  & 50,147 & 2,008 & 2,051 & 17,720 \\
        \bottomrule
    \end{tabular}
\end{table}

Preliminary findings show significant variation in the synchrony between code and documentation updates across projects, fluctuating over time due to various factors,
which our study aims to identify. Table~\ref{tab:repo_stats} shows the quantitative analysis results, significant differences between Code, Documentation, and Configuration churns are reported. Mix represents our interest where there is visible linkage between code and other updates, however, it is still comparatively lower than code churns. For instance, HTTP project has 14,768 code churns over time, which were never linked to documentation, and only 5,417 were synchronized.
In all three projects, documentation is significantly smaller (2.595, 1.893, 2.008) compared to code churns (14,768; 22,317; 50,147), indicating a substantial knowledge gap. Our quantitative analysis revealed that documentation updates were not continuously linked to code churns over time, even when new features were introduced to the codebase. Periods of code changes often lacked corresponding documentation updates, while mixed churns facilitated smoother knowledge transfer within the community.
\section{Immediate Future Work with LLMs}
\label{discussion}

Findings highlight variability in documentation practices, showing that documentation is sometimes linked to code churn changes but not always when necessary. This reveals a need for improved strategies, possibly using current LLM models and automatic documentation generation, for better knowledge transfer.

Our future work will expand the dataset to include projects like PyPI, NPM, and GNU libraries, analyse developer activity over time, and study the correlation between open and closed issues and churn categories. 
We look to understand why certain groups of contributors neglect documentation updates and what the different situations are where this applies. 
Also, in what cases AI-powered approaches can suggest new strategies for software maintenance.
There may be cases where the documentation might not be needed.

We will focus on understanding the causes of variability and finding effective strategies to improve documentation practices in open-source communities by extending analysis into other documentation such as GitHub issues and Pull Requests. These artefacts may reveal additional insights. Other key areas include exploring advanced techniques like semantic analysis for aligning code changes with documentation updates and examining contextual factors influencing synchronization.

\section{Acknowledgments}

This work is supported by JST BOOST JPMJBS2423, JSPS KAKENHI JP20H05706, 	JP23K28065.

\bibliographystyle{IEEEtran}
\bibliography{bibliography}

\begin{thebibliography}{10}
\providecommand{\url}[1]{#1}
\csname url@samestyle\endcsname
\providecommand{\newblock}{\relax}
\providecommand{\bibinfo}[2]{#2}
\providecommand{\BIBentrySTDinterwordspacing}{\spaceskip=0pt\relax}
\providecommand{\BIBentryALTinterwordstretchfactor}{4}
\providecommand{\BIBentryALTinterwordspacing}{\spaceskip=\fontdimen2\font plus
\BIBentryALTinterwordstretchfactor\fontdimen3\font minus \fontdimen4\font\relax}
\providecommand{\BIBforeignlanguage}[2]{{%
\expandafter\ifx\csname l@#1\endcsname\relax
\typeout{** WARNING: IEEEtran.bst: No hyphenation pattern has been}%
\typeout{** loaded for the language `#1'. Using the pattern for}%
\typeout{** the default language instead.}%
\else
\language=\csname l@#1\endcsname
\fi
#2}}
\providecommand{\BIBdecl}{\relax}
\BIBdecl

\bibitem{gaikovina2023life}
R.~G. Kula and G.~Robles, ``The life and death of software ecosystems,'' in \emph{Towards Engineering Free/Libre Open Source Software (FLOSS) Ecosystems for Impact and Sustainability: Communications of NII Shonan Meetings}.\hskip 1em plus 0.5em minus 0.4em\relax Springer, 2019, pp. 97--105.

\bibitem{favre2012linking}
J.-M. Favre, R.~L{\"a}mmel, M.~Leinberger, T.~Schmorleiz, and A.~Varanovich, ``Linking documentation and source code in a software chrestomathy,'' in \emph{2012 19th Working Conference on Reverse Engineering}.\hskip 1em plus 0.5em minus 0.4em\relax IEEE, 2012, pp. 335--344.

\bibitem{rai22review}
S.~Rai, R.~C. Belwal, and A.~Gupta, ``A review on source code documentation,'' \emph{ACM Transactions on Intelligent Systems and Technology (TIST)}, vol.~13, no.~5, pp. 1--44, 2022.

\bibitem{mcburney14auto}
P.~W. McBurney, ``Automatic documentation generation via source code summarization,'' in \emph{2015 IEEE/ACM 37th IEEE International Conference on Software Engineering}, vol.~2.\hskip 1em plus 0.5em minus 0.4em\relax IEEE, 2015, pp. 903--906.

\bibitem{shameer2023relationship}
S.~Shameer, G.~Rodr{\'\i}guez-P{\'e}rez, and M.~Nagappan, ``Relationship between diversity of collaborative group members’ race and ethnicity and the frequency of their collaborative contributions in github,'' \emph{Empirical Software Engineering}, vol.~28, no.~4, p.~83, 2023.

\bibitem{davidson2014older}
J.~L. Davidson, R.~Naik, U.~A. Mannan, A.~Azarbakht, and C.~Jensen, ``On older adults in free/open source software: reflections of contributors and community leaders,'' in \emph{2014 IEEE Symposium on Visual Languages and Human-Centric Computing (VL/HCC)}.\hskip 1em plus 0.5em minus 0.4em\relax IEEE, 2014, pp. 93--100.

\bibitem{robles2016women}
G.~Robles, L.~A. Reina, J.~M. Gonz{\'a}lez-Barahona, and S.~D. Dom{\'\i}nguez, ``Women in free/libre/open source software: The situation in the 2010s,'' in \emph{Open Source Systems: Integrating Communities: 12th IFIP WG 2.13 International Conference, OSS 2016, Gothenburg, Sweden, May 30-June 2, 2016, Proceedings 12}.\hskip 1em plus 0.5em minus 0.4em\relax Springer, 2016, pp. 163--173.

\bibitem{de2024fine}
C.~R. de~Souza, G.~Rodr{\'\i}guez-P{\'e}rez, M.~Basha, D.~Yoon, and I.~Beschastnikh, ``The fine balance between helping with your job and taking it: Ai code assistants come to the fore,'' \emph{IEEE Software}, 2024.

\bibitem{white2024chatgpt}
J.~White, S.~Hays, Q.~Fu, J.~Spencer-Smith, and D.~C. Schmidt, ``Chatgpt prompt patterns for improving code quality, refactoring, requirements elicitation, and software design,'' in \emph{Generative AI for Effective Software Development}.\hskip 1em plus 0.5em minus 0.4em\relax Springer, 2024, pp. 71--108.

\bibitem{robles2024role}
G.~Robles, C.~Treude, J.~M. Gonzalez-Barahona, and R.~G. Kula, ``The role of code proficiency in the era of generative ai,'' \emph{arXiv preprint arXiv:2405.01565}, 2024.

\end{thebibliography}

\end{document}